\documentclass[aps,prb,epsfig,twocolumn]{revtex4}

\begin{document}

\bibliographystyle{prsty}

\draft

\title{Limit Cycle and Conserved Dynamics}
\author{ X.-M. Zhu$^{1)}$, L. Yin$^{2)}$, and P. Ao$^{3)}$ }
\address{ $^{1)}$GenMath, 5525 27th Ave. N.E., Seattle, WA 98105, USA  \\
                $^{2)}$School of Physics, Peking University, 100871 Beijing, PR China \\
                $^{3)}$Department of Mechanical Engineering,
                    University of Washington, Seattle, WA 98195, USA  }
\date{ Dec. 7, 2004 }


\begin{abstract}
We demonstrate that a potential coexists with limit cycle. Here
the potential determines the final distribution of population. Our
demonstration consists of three steps: We first show the existence
of limit from a typical physical sciences setting: the potential
is a type of Mexican hat type, with the strength of a magnetic
field scale with the strength the potential gradient near the
limit cycle, and the friction goes to zero faster than the
potential near the limit cycle. Hence the dynamics at the limit
cycle is conserved. The diffusion matrix is nevertheless finite at
the limit cycle. Secondly, we construct the potential in the
dynamics with limit cycle in a typical dynamical systems setting.
Thirdly, we argue that such a construction can be carried out in a
more general situation based on a method discovered by one of us.
This method of dealing with stochastic differential equation is in
general different from both Ito and Stratonovich calculus. Our
result may be useful in many related applications, such as in the
discussion of metastability of limit cycle and in the construction
of Hopfield potential in the neural network computation.
\end{abstract}


\maketitle

\section{Introduction}

Noise perturbed dynamics systems with limit cycles are abounded in
physical sciences. It is one of the driving forces in the study of
nonlinear systems \cite{lindner,ao2004a}. Until recently, one of
the most useful concepts in physical sciences, the potential
energy function, had been found to be not applicable in general
\cite{ao2004a,guckenheimer}. Hence, it has been concluded that the
no Lyapunov function or potential function with periodic
attractors, in the neural network computing \cite{hertz}. Similar
statements can also be found in many other fields. When the
matastability and the decay of metability state are discussed,
efforts have been made to avoid the use of potential energy
function in the form of Boltzmann-Gibbs distribution. Various
methods which can go around this issue, such as the
Machlup-Onsager functional method, quasipotentials, etc, have been
developed \cite{berglund}. They have been successfully applied to
situations with limit cycle dynamics, such as in the discussion of
topological structure of the decay of metastable limit cycle
\cite{dykman}, in the numerical study of trajectory of the
escaping path \cite{mcclintock}. However, even in such settings,
there is a need to consider potential energy function, such as
indicated in the escape of asteroids in the solar system
\cite{jaffe}. Hence, it is a natural question to ask what would be
the connection between such an approach and the classical approach
based on potential energy function \cite{kramers,hanggi}, and
whether or not the potential function exists in dynamics with
limit cycle. The purpose of the present article is to give an
explicit demonstration that at least for the case of limit cycle
dynamics, potential function can be explicitly constructed, which
provides a critical link between those two ways of dealing with
metastability.

Our construction is based on a novel way of handling the
stochastic differential equation. It can be understood as the zero
mass limit of a generalized Klein-Kramers equation \cite{ao2004a}.
In this limit the steady state distribution can be established
from the Klein-Kramers equation without the concerns of the Ito or
Stratnovich calculus. In this sense it would not be surprising
that a different perspective can be obtained other than those in
Ref.[\cite{dykman,mcclintock,jaffe}]. We believe that our
following demonstration provides an important step for an
alternative understanding of the nonequilibrium processes such as
limit cycles.

We will demonstrate the co-existence of a potential with limit
cycle in three steps in three sections, respectively. In section
II we explicitly demonstrate how to construct a limit cycle based
on our knowledge in physical sciences, with a potential. This is
different from the usual approach in dynamical systems. In section
III we demonstrate how to construct a potential in the presence of
a limit cycle in the dynamical systems setting. In section IV we
outline a general construction of potential in a broad class of
dynamical systems, including limit cycles.  In section V we
discuss two mathematical subtleties. In section VI we conclude
that the potential can co-exist with limit cycle and with possibly
more complicated dynamics.

\section{Limit cycle: physical sciences` point of view}

In physical sciences, the general dynamical equation for a {\it
massless} particle in two dimensional state space may be expressed
as, with both deterministic and stochastic forces
\cite{goldstein}:
\begin{equation}
  [ S({\bf q}, t) + T ({\bf q}, t) ] \dot{\bf q}
          =  -  \nabla \psi ({\bf q}, t) + {\bf \xi}({\bf q}, t) \; ,
\end{equation}
and supplemented by the relationship on the stochastic force:
\begin{equation}
  \langle {\bf \xi}({\bf q}, t) {\bf \xi}^{\tau} ({\bf q}, t') \rangle
         = 2 S({\bf q}, t) \; \epsilon  \; \delta( t-t' ) \; ,
\end{equation}
and $\langle {\bf \xi}({\bf q}, t) \rangle = 0$. Here ${\bf
q}^{\tau} = (q_1, q_2)$ with $q_1, q_2$ the two Cartesian
coordinates of the state space, which may be perceived as the
position space of the massless particle. The transpose is denoted
by the superscript $\tau$, and $\dot{\bf q}  = d {\bf q}/dt $.

The scalar function $\psi$ will is the usual potential energy
function. Its graphical representation in the state space is a
landscape. The antisymmetric matrix $T$ represents the dynamics
which conserves the potential, corresponding to the Lorentz force
in physical sciences, determined by the magnetic field. The matrix
$S$ represents the dynamics which decreases the potential, the
dissipation. This matrix will be called friction matrix. The
friction matrix is connected to the stochastic force ${\bf \xi}$
by Eq.(2), which guarantees that it is nonnegative and symmetric.
All $T, S, \psi$ can be nonlinear functions of the state variable
${\bf q}$ as well as the time $t$. The numerical parameter
$\epsilon$ corresponds to an effective temperature, which will be
taken to be zero to recover the deterministic dynamics. It has
been shown that if a steady state distribution $\rho({\bf q})$ in
state space exists,
\begin{equation}
    \rho({\bf q})  \propto \exp
        \left( - \frac{\psi ({\bf q}) }{\epsilon } \right) \; ,
\end{equation}
a Boltzmann-Gibbs type distribution function \cite{ao2004a}.
Eq.(3) implies that for dynamics which repeats itself
indefinitely, such as limit cycle, the potential may be the same
along such trajectory.

It should be pointed that the potential function $\psi ({\bf q},
t)$ exists from the beginning by construction. This is one of most
useful quantitative concepts in physical sciences. If the
stochastic force could be set to be zero, ${\bf \xi}({\bf q}, t) =
0$, that is, the deterministic dynamics, the dynamics of this
massless particle always decreases its potential energy:
\begin{equation}
 \begin{array}{lcl}
   \dot{\bf q}^{\tau} \nabla \psi ({\bf q}, t)
    & = & - \dot{\bf q}^{\tau} [ S({\bf q}, t) + T ({\bf q}, t) ]
            \dot{\bf q}   \\
    & = & - \dot{\bf q}^{\tau} S({\bf q}, t) \dot{\bf q}  \\
    & \leq &  0 \; .
 \end{array}
\end{equation}
The zero occurs only at the limiting set. Hence the potential
function has the usual meaning of Lyapunov function.

To model a limit cycle, we choose following forms for the friction
matrix $S$, the anti-symmetric matrix $T$, and the potential
$\psi$, assuming the limit cycle occurs at $q = \sqrt{q_1^2  +
q_2^2 } = 1$:
\begin{eqnarray}
   S & = & \frac{ (q^2 -1)^2 }{ (q^2 - 1)^2 + 1 }
                           \left( \begin{array}{ll}
                                      1 & 0  \\
                                      0 & 1
                           \end{array} \right)   \\
   T &  = & (q-1) \frac{ q^2 }{ (q^2 - 1 )^2 + 1 }
                                 \left( \begin{array}{ll}
                                          0 & -1  \\
                                         1 & 0
                                  \end{array} \right)    \\
   \psi & = &   \frac{1}{2} (q - 1)^2
\end{eqnarray}
The potential $\psi$ given in Eq.(7) is rotational symmetric in
the state space. It has a local maximum $\psi = 1/2$ at $q=0$,
which is a fixed point, and the minimum $\psi = 0$ at $q=1$, which
is a cycle in the state space. Hence the potential takes the shape
of Mexican hat type.

Evidently, if the friction matrix $S$ would be zero, the dynamical
trajectory of the massless particle would move along the equal
potential contour determined by the initial condition, which would
be a cycle according to above chosen potential. In the presence of
nonzero friction matrix, this is not true. What will be our
concern is the behavior near the minimum of the potential
function: When $q$ is sufficiently close to 1, does the particle
trajectory asymptotically approach the cycle of $q=1$ and
eventually coincide with it? If the answer is positive, we have a
limit cycle dynamics. We will demonstrate below that it is indeed
possible.

For a deterministic dynamics, we can set $\epsilon = 0$ in Eq.(1)
and (2): setting the stochastic force to be zero. The dynamical
equation can then be rewritten as
\begin{equation}
      \dot{\bf q}
          =  - [ S({\bf q},t) + T ({\bf q}, t) ]^{-1}
             \nabla \psi ({\bf q}, t)
\end{equation}
With the choice of Eqs. (5-7), we have
\begin{eqnarray}
  [ S + T ]^{-1}
   &  =  & \frac{1}{\det(S + T ) }
       \left[  \frac{ (q^2 -1)^2 }{ (q^2 - 1)^2 + 1 }
        \left( \begin{array}{ll}
                 1 & 0  \\
                 0 & 1
        \end{array} \right)               \right. \nonumber  \\
  &  &  \left. - (q-1) \frac{ q^2 }{ (q^2 - 1 )^2 + 1 }  )
            \left( \begin{array}{ll}
                 0 & -1  \\
                 1 & 0
            \end{array} \right)     \right]
\end{eqnarray}
and $\det(S+T) = [(q^2 - 1)^2 /((q^2 - 1)^2 + 1)]^2
                          +  [(q - 1)^2  q^4  /[(q^2 - 1)^2 + 1 ] ^2 $.
Near $q=1$, we have
\begin{eqnarray}
  [ S + T ]^{-1} & =  & \frac{1}{q-1} \left[ - (1-  2  (q-1)  )
           \left( \begin{array}{ll}
                 0 & -1  \\
                 1 & 0
           \end{array} \right)  \right.  \nonumber  \\
           &   &    + \left.  4 (q-1)
            \left( \begin{array}{ll}
                 0 & -1  \\
                 1 & 0
            \end{array} \right)     + O( (q-1)^2 )  \right]
\end{eqnarray}
In terms of radial coordinate $q$ and azimuth angle $\theta$
%
%
in the polar coordinate representation of the state space,  using
the small parameter expansion given in Eq.(10) following Eq.(8) we
have, to the order of $q-1$,
\begin{eqnarray}
   \dot{q}&  = & -  4  (q-1)  \\
   \dot{\theta }&  = & 1 -  2 \frac{q-1 }{q}
\end{eqnarray}
The solution is
\begin{eqnarray}
  q(t) &  =  & 1 + \delta q_0 \exp\left\{ -  4  t  \right \} \\
   \theta(t) & = & \theta_0 + t
       + \frac{1}{4} \ln \frac{1+ \delta q_0
          \exp\left\{ - 4  t \right\} }
           {1+ \delta q_0  }  \nonumber  \\
    &  &  + \frac{1}{2} \ln\left(1+ \delta q_0
      \exp\left\{-  4  t \right\} \right)
\end{eqnarray}
Here $\delta q_0$ ($| \delta q_0 | << 1$) is the starting radial
position of the particle measured from $q=1$ and $\theta_0$ its
the starting azimuth angle. Indeed, the solution demonstrates the
asymptotical approaching to the cycle $q=1$, and the motion never
stops. Unstable and metable limit cycles may be constructed in the
similar manner.

Though above construction does show that based on the knowledge of
physical sciences one can construct limit cycle with a potential,
the example of Eqs.(5-7) appears contrived. We should, however,
point out that there are several generic features in our
construction. First, to have an indefinitely motion on a closed
trajectory, because of energy conservation, the friction, or
better the friction matrix here, must be zero. Second, because of
the asymptotic motion is on the equal potential contour, a Lorentz
force type must exist to keep the motion on the contour. This
means that the antisymmetric matrix should be finite along the
equal motion contour when the potential gradient is finite. The
speed of the massless particle moving along the contour will be
determined by the ratio of the strength of the Lorentz like force
to that of the gradient of the potential. Third, because the limit
cycle should be robust: Small parameter changes should only have a
small effect on the limit cycle, the (stable) limit cycle should
be at the minimum of the potential. As a consequence, the
potential gradient at the maximum is zero, which would imply that
the friction matrix must go to zero faster than that of the
potential gradient when approaching to the maximum to avoid the
potential taking singular values. This means that at the limit
cycle the dynamics is conserved. All those features are explicitly
implemented in the choice, Eqs.(5-7).

Two additional remarks are in order.  For simplicity of
calculation we have chosen the friction matrix to be a constant
matrix in Eq.(5). One can check that any positive definite
symmetric matrix can lead to above same conclusion, as long as its
strength goes to zero in a higher order comparing to that of the
potential gradient. The second remark is that although $S$ is zero
when approaching to the limit cycle, $[S + T]^{-1} + [S - T
]^{-1}$ is not, on which we will come back later when discussing
the diffusion matrix in section V.

\section{Limit cycle: simple case in dynamical systems}

The demonstration of the co-existence of limit cycle with
potential in section II may appear special: The particle moves
along the potential minimum with both zero friction and zero
transverse matrices, a nice picture from physical sciences. A
question naturally arises that whether or not in a typical limit
cycle in dynamical systems a potential can be constructed.  We
demonstrate in this section that the answer is a straightforward
yes.

It has been suggested that a typical and simple limit cycle in two
dimensions would take following dynamical equation in a polar
coordinate representation \cite{murray}:
\begin{eqnarray}
   \dot{q }      &  = &  R(q)   \\
   \dot{\theta} & = &  \Phi (q)
\end{eqnarray}
Here the smooth functions $R,\Phi$ have properties $R(q =  1) = 0
$ is a fixed point in the radial coordinate and $\Phi (q=1) =
constant$. It is mathematically possible that any limit cycle in
two dimension can be deformed into a cycle, and, at least near
this limit cycle, the dynamical equation can be mapped into above
form by a nonlinear coordinate transformation.

The comparison between Eqs.(11,12) and Eqs.(15,16) immediately
suggests what considered in section II is just such a typical
limit cycle.  Hence, we can conclude that the potential and limit
cycle coexist for such cases.

Nevertheless, we should point out an interesting paradoxical
feature. On one side it is known from the theory of dynamical
systems that a limit cycle is robust \cite{guckenheimer}. Any
small parameter change in the equation would not lead to its
disappearing.  On other side, from the demonstration in section II
which is based on the understanding from physical sciences, the
existence of the limit cycle is a result of a very delicate
balance between all dynamical elements: The friction matrix, the
gradient of potential function, and the antisymmetric matrix are
all zero at the limit cycle. How this feature would play a role in
our better understanding limit cycle and its control will be an
interesting problem for further exploration.

\section{General construction of evolution potential}

A particular challenging and careful reader would observe that we
have not demonstrated the potential for a general limit cycle in
arbitrary dimensions. Hence there might exist a type of limit
cycle, which will not be ones such as discussed above, such that a
potential might not exist. We have been unable to give an explicit
demonstration for the existence of potential for a most generic
limit cycle (periodic attractors). However, we point out that such
a discussion has been done for a generic fixed point (point
attractors) \cite{kat}. Furthermore, we do have a generic
construction of potential for a general situation. This
construction suggests it is possible to do so generally for
arbitrary limit cycles. The general method has been presented
elsewhere \cite{ao2004a}. Here we outline its key ideas to make
the present demonstration complete.

We restore the stochastic force back into the general equation in
dynamical systems and consider an arbitrary dimension $n$ instead
merely two dimensions:
\begin{equation}
  \dot{\bf q} = {\bf f}({\bf q}, t) +  {\bf \zeta}( {\bf q} , t) \; .
\end{equation}
Here ${\bf f}({\bf q}, t)$ is the deterministic nonlinear drive of
the system and the stochastic drive is ${\bf \zeta}( {\bf q}, t)$,
which differs from that in Eq.(1) but comes from the same
dynamical source. For simplicity we will assume that ${\bf f}$ is
a smooth function whenever needed. To be consistent with Eq.(1)
and (2), the stochastic drive in Eq.(17) is also assumed to be
Gaussian and white with the variance,
\begin{equation}
 \langle {\bf \zeta}( {\bf q}, t)
         {\bf \zeta}^{\tau}({\bf q}, t') \rangle
     = 2 D( {\bf q}, t) \; \epsilon \; \delta (t-t') ,
\end{equation}
and zero mean, $\langle {\bf \zeta}({\bf q}, t)\rangle = 0$. The
matrix $D$ is the diffusion matrix.

Assuming that both Eq.(1) and (17) describe the same dynamics. It
is rather easy to derive Eq.(17) from Eq.(1). The difficult part
is to derive Eq.(1) from Eq.(17). Using Eq.(17) to eliminate
$\dot{\bf q}$ in Eq.(1), and noticing that the dynamics of noise
and the state variable behave independently, we have, the
deterministic part,
\begin{equation}
  [S({\bf q}, t) + T({\bf q}, t)] {\bf f}({\bf q}, t)
      = - \nabla \psi({\bf q}, t) \; ,
\end{equation}
and the stochastic part,
\begin{equation}
   [S({\bf q}, t) + T({\bf q}, t)] {\bf \zeta}({\bf q}, t)
       = \xi({\bf q}, t) \; .
\end{equation}
Those two equations suggest a rotation in state space.

Multiplying Eq.(20) by its own transpose of each side and carrying
out the average over stochastic drive, we have
\begin{equation}
  [S({\bf q}, t) + T({\bf q}, t) ] D({\bf q}, t)
  [S({\bf q}, t) - T({\bf q}, t) ] = S({\bf q}, t) \; .
\end{equation}
In obtaining Eq.(21) we have also used Eq.(2) and (18). Eq.(21)
gives $n(n+1)/2$ conditions because it is symmetric under the
transpose operation.

Using the property of the potential $\psi$: $\nabla \times \nabla
\psi = 0$ [$(\nabla \times \nabla \psi)_{ij} = (\nabla_i \nabla_j
-\nabla_i \nabla_j)\psi $ ],   Eq.(19) leads to
\begin{equation}
  \nabla \times [  [S({\bf q}, t) + T({\bf q}, t) ]
          {\bf f}({\bf q}) ] = 0 \; ,
\end{equation}
which gives $n(n-1)/2$ conditions because the antisymmetric
condition implied.

Combining both Eq.(21) and (22), we have total $n^2$ conditions,
which is exactly the number to determine the $n\times n$ matrix
$[S({\bf q}, t) + T({\bf q}, t) ]$, when supplemented by
appropriate boundary condition needed by the partial differential
equations of Eq.(22). This additional boundary condition should be
the delicate balance condition discussed in section II and III:
When approaching the limit cycle, the friction matrix must
approach to zero faster than the potential gradient as well as the
antisymmetric matrix. Once $[S + T ]$ is known, it is
straightforward to obtain the potential according to Eq.(19),
hence construct Eq.(1) and (2) from Eq.(17) and (18).

The general existence of the potential implies that indeed the
energy landscape concept in physical sciences should be one of the
most important concepts in dynamics systems, too. Thus, not only
the potential exist in periodic attractors, may also in the
strange attractors. This would have implications in other fields
where dynamical behaviors are the primary concerns. For example,
we have shown elsewhere \cite{ao2004b} that the mathematical
structure of population genetics can be formulated in such a way
to incorporate the important concept of potential.

\section{Discussions}

There are two mathematical subtleties. First, it is known that
even with a limit cycle the dynamics is dissipative, reflecting by
the fact that in general $\nabla \cdot {\bf f} \neq 0 $, where $f$
is defined by Eq.(17). This is also expressed by the fact that the
so-called diffusion matrix, $D$ defined in Eq.(18), is finite even
at the limit cycle. Because it is dissipative, it would be
difficult to conceive a constant potential (or a Lyapunov
function) along the limit cycle on which the dynamics repeats
itself indefinitely. The delicate point is that, as shown in our
above demonstration, that the friction matrix $S$ is zero along
the limit cycle does not implies the diffusion matrix $D$ is zero.
In fact, it is finite according to Eq.(21):
\begin{equation}
   D = \frac{1}{2} [ (S + T)^{-1} + (S - T)^{-1} ]  \;  .
\end{equation}
An interesting and direct implication is that the statement that
one cannot construct Hopfield potential (or Lyapunov function) in
limit cycle in computing by attractors in the neural networks
\cite{hertz} is not true.

We come to the second point. For deterministic dynamics, if one
finds one Lyaponov function, one finds many. This is illustrated
by the present construction that additional information from the
noise is needed: different diffusion matrix would lead to
different potential function. Hence, this may provide a method to
select best suitable Lyapunov function or potential function to
one's own problem by choosing appropriate form the diffusion
matrix.

It is perhaps worthwhile to point out that the typical gradient
systems in dynamical systems theory corresponds to the zero
transverse matrix, $T=0$, in the present construction. No limit
cycle is possible in this case. In dynamics described by gradient
systems the trajectory could follow the most rapid descendant path
along the landscape defined by the potential. In this case it is
easy to identify the potential as the landscape function.  For a
general dynamics where the transverse matrix is not zero, the
trajectory would not follow the most rapid descendant route along
the potential, as expressed by Eq.(1). Nevertheless, the meaning
of the potential remains the same as that in gradient systems:
determining the final distribution.

\section{Conclusions}

Based on the demonstrations in section II-IV, we conclude that a
potential can coexists with limit cycle. Such a potential
determines the final distribution according to a Boltzmann-Gibbs
distribution. This may provide an alternative perspective to study
the metastability of limit cycle from Kramers' potential landscape
point of view. In regarding to computing by attractors in neural
networks, the present article shows that the Hopfield potential
exists for periodic attractors.

{\ }

\noindent{\bf Acknowledgement:} We thank M. Dykman and H. Qian for
critical discussions. This work was supported in part by a USA NIH
grant under HG002894.

{\ }

\end{document}